# Solution of non-singlet DGLAP evolution equation in leading and next-to-leading order at small-x by method of characteristics

R. Baishya[1]

Physics Department, J.N. College, Boko-781123, Assam, India

J. K. Sarma[2]

Physics Department, Tezpur University, Napam-784028, Assam, India

The non-singlet structure functions have been obtained by solving Dokshitzer-Gribov-Lipatov-Alterelli-Parisi (DGLAP) evolution equations in leading order (LO) and next-to-leading order (NLO) at the small-x limit. Here a Taylor series expansion has been used and then the method of characteristics has been applied to solve the evolution equations. Results are compared with the Fermilab experiment E665 data and New Muon Collaboration (NMC) data.

In our recent report [1], we have explained the importance of DGLAP evolution equations and their solutions, and also the importance of the method of characteristics to solve those equations. There, we have solved singlet DGLAP evolution equation in LO and NLO at small-x [2] by using the method of characteristics and compared deuteron structure function thereby obtained with NMC data with satisfactory phenomenological success. In this brief report, we have solved non-singlet DGLAP evolution equation in LO and NLO at small-x by using the same method and compared non-singlet structure function thereby obtained with Fermilab experiment E665 data [3] and NMC data [4] with considerable phenomenological success. The non-singlet structure function is the combination of only valence quark densities and free of singlet quark and gluon densities. Therefore when solving the non-singlet DGLAP equation, we need not relate quark and gluon structure functions by a suitable function k(x) or by a constant as in the case of solution of the singlet DGLAP evolution equation given in our earlier report [1]. So, the solution of non-singlet equation is a parameter free solution and this is a more stringent test of our method in solving DGLAP evolution equations.

[1] E-mail:- rjitboko@yahoo.co.in
[2] E-mail :- jks@tezu.ernet.in

The DGLAP evolution equation in the non singlet sector in the standard form [5-7] is given by

$$\frac{\partial F_2^{NS}\left(x, Q^2\right)}{\partial \ln Q^2} = P_{NS}\left(x, Q^2\right) \otimes F_2^{NS}\left(x, Q^2\right), \tag{1}$$

where $F_2^{NS}\left(x, Q^2\right)$ is the non-singlet structure function as the function of x and $Q^2$, where x is the Bjorken variable and $Q^2$ is the four momentum transfer in a DIS process. Here $P_{NS}(x, Q^2)$ is the non-singlet kernel known pertarbatively up to the first few orders in $\alpha_S\left(Q^2\right)$, the strong coupling constant. With the notation

$$a(x) \otimes b(x) \equiv \int_0^1 \frac{dy}{y} a(y) b\left(\frac{x}{y}\right), \tag{2}$$

we can write

$$P_{NS}\left(x, Q^2\right) = \frac{\alpha_S\left(Q^2\right)}{2\pi} P^{(0)}{}_{NS}(x) + \left(\frac{\alpha_S\left(Q^2\right)}{2\pi}\right)^2 P^{(1)}{}_{NS}(x) + \Lambda\ \Lambda\ , \tag{3}$$

where $P^{(0)}{}_{NS}(x)$, $P^{(1)}{}_{NS}(x)$, ⋯⋯ are non-singlet splitting functions in LO, NLO, ⋯⋯ respectively.

Appling all these and simplifying, the DGLAP evolution equations for non-singlet structure function in LO and NLO can be written as [6]

$$\frac{\partial F_2^{NS}}{\partial t} - \frac{\alpha_S(t)}{2\pi}\left[\frac{2}{3}\{3 + 4\ln(1-x)\} F_2^{NS}(x, t) + I_1^{NS}(x, t)\right] = 0 \tag{4}$$

and

$$\frac{\partial F_2^{NS}}{\partial t} - \frac{\alpha_S(t)}{2\pi}\left[\frac{2}{3}\{3 + 4\ln(1-x)\} F_2^{NS}(x, t) + I_1^{NS}(x, t)\right] - \left(\frac{\alpha_S(t)}{2\pi}\right)^2 I_2^{NS}(x, t) = 0 \tag{5}$$

respectively, where

$$I_1^{NS}(x, t) = \frac{4}{3}\int_x^1 \frac{d\omega}{1-\omega}\left[\left(1+\omega^2\right) F_2^{NS}\left(\frac{x}{\omega}, t\right) - 2F_2^{NS}(x, t)\right], \tag{6}$$

$$I_2^{NS}(x, t) = (x-1) F_2^{NS}(x, t) \int_0^1 f(\omega) d\omega + \int_x^1 f(\omega) F_2^{NS}\left(\frac{x}{\omega}, t\right) d\omega \tag{7}$$

and $t = \ln \frac{Q^2}{\wedge^2}$, where $\wedge$ is the QCD cut off parameter. Here

$$f(\omega) = C_F^2\left[P_F(\omega) - P_A(\omega)\right] + \frac{1}{2}C_F C_A\left[P_G + P_A(\omega)\right] + C_F T_R N_f P_{N_f}(\omega),$$

$$P_F(\omega) = -\frac{2(1+\omega^2)}{(1-\omega)}\ln(\omega)\ln(1-\omega) - \left(\frac{3}{1-\omega} + 2\omega\right)\ln\omega - \frac{1}{2}(1+\omega)\ln\omega + \frac{40}{3}(1-\omega),$$

$$P_G(\omega) = \frac{(1+\omega^2)}{(1-\omega)}\left(\ln^2(\omega) + \frac{11}{3}\ln(\omega) + \frac{67}{9} - \frac{\pi^2}{3}\right) + 2(1+\omega)\ln\omega + \frac{40}{3}(1-\omega),$$

$$P_{N_f}(\omega) = \frac{2}{3}\left[\frac{1+\omega^2}{1-\omega}\left(-\ln\omega - \frac{5}{3}\right) - 2(1-\omega)\right]$$

and

$$P_A(\omega) = \frac{2(1+\omega^2)}{(1+\omega)}\int_{\left(\frac{\omega}{1+\omega}\right)}^{\left(\frac{1}{1+\omega}\right)} \frac{dk}{k}\ln\left(\frac{1-k}{k}\right) + 2(1+\omega)\ln(\omega) + 4(1-\omega),$$

with $C_A = C_G = 3$, $C_F(\omega) = \frac{N_f^2 - 1}{2N_f}$, $T_f(\omega) = \frac{1}{2}$ and $\alpha_S(t) = \frac{4\pi}{\beta_0 t}\left[1 - \frac{\beta_1 \ln t}{\beta_0^2 t}\right]$. $\beta_0 = 11 - \frac{2}{3}N_f$ and $\beta_1 = 102 - \frac{38}{3}N_f$ are the one loop (LO) and two loop (NLO) corrections to the QCD $\beta$-function and $N_f$ being the flavour number. We can neglect $\beta_1$ for LO.

Let us introduce the variable $u = 1-\omega$ and note that $\frac{x}{\omega} = \frac{x}{1-u} = x + \frac{xu}{1-u}$. Since $x<\omega<1$, so $0<u<1-x$, and hence the series is convergent for $|u|<1$. Now using Taylor's expansion series, we can rewrite

$$F_2^{NS}\left(\frac{x}{\omega}, t\right) = F_2^{NS}\left(x + \frac{xu}{1-u}, t\right)$$

$$= F_2^{NS}(x,t) + \frac{xu}{1-u}\frac{\partial F_2^{NS}(x,t)}{\partial x} + \frac{1}{2}\left(\frac{xu}{1-u}\right)^2 \frac{\partial^2 F_2^{NS}(x,t)}{\partial x^2} + \Lambda\ \Lambda \quad .$$

Since x is small in our region of discussion, the terms containing $x^2$ and higher powers of x can be neglected. So we get

$$F_2^{NS}\left(\frac{x}{\omega}, t\right) = F_2^{NS}(x,t) + \frac{xu}{1-u}\frac{\partial F_2^{NS}(x,t)}{\partial x} \quad . \qquad (8)$$

Using equation (8) in equation (6) and performing u-integrations, equation (4) becomes the form

$$-t\frac{\partial F_2^{NS}(x,t)}{\partial t}+\frac{3}{2}A_f A(x)F_2^{NS}(x,t)+\frac{3}{2}A_f B(x)\frac{\partial F_2^{NS}(x,t)}{\partial x}=0, \quad (9)$$

where $A_f=\frac{\alpha_S(t)}{3\pi}t=\frac{4}{3\beta_0}=\frac{4}{33-2N_f}$, $A(x)=2x+x^2+4\ln(1-x)$ and

$B(x)=x-x^3-2x\ln(x)$.

To introduce the method of characteristics, let us consider two new variables S and $\tau$ instead of x and t, such that $\frac{dt}{dS}=-t$ and $\frac{dx}{dS}=A_f B(x)$, which are known as characteristic equations. Putting these in equation (9), we get

$$\frac{dF_2^{NS}(S,\tau)}{dS}+L(S,\tau)F_2^{NS}(S,\tau)=0,$$

which can be solved as

$$F_2^{NS}(S,\tau)=F_2^{NS}(\tau)\left(\frac{t}{t_0}\right)^{L(S,\tau)},$$

where $L(S,\tau)=\frac{3}{2}A_f A(x)$ and $F_2^{NS}(S,\tau)=F_2^{NS}(\tau)$ for initial condition $S=0\Rightarrow t=t_0$. Now we have to replace the co-ordinate system (S, $\tau$) to (x, t) with the input function $F_2^{NS}(0,\tau)=F_2^{NS}(x,t_0)$ and will get the t-evolution of non-singlet structure function in LO as

$$F_2^{NS}(x,t)=F_2^{NS}(x,t_0)\left(\frac{t}{t_0}\right)^{\frac{3}{2}A_f A(x)}. \quad (10)$$

Similarly the x-evolution of non-singlet structure function in LO will be

$$F_2^{NS}(x,t)=F_2^{NS}(x_0,t)\exp\left[-\int_{x_0}^{x}\frac{A(x)}{B(x)}dx\right]. \quad (11)$$

In NLO, the t and x evolutions of non-singlet structure functions will be

$$F_2^{NS}(x,t)=F_2^{NS}(x,t_0)\left(\frac{t}{t_0}\right)^{\frac{3}{2}A_f[A(x)+T_0A_1(x)]} \quad (12)$$

and

$$F_2^{NS}(x,t)=F_2^{NS}(x_0,t)\exp\left[-\int_{x_0}^{x}\frac{A(x)+T_0A_1(x)}{B(x)+T_0B_1(x)}dx\right] \quad (13)$$

respectively, with $A_1(x)=x\int_0^1 f(\omega)d\omega-\int_0^x f(\omega)d\omega+\frac{4}{3}N_f\int_x^1 F_{qq}(\omega)d\omega$ and

$B_1(x)=x\int_x^1\left[f(\omega)+\frac{4}{3}N_f F_{qg}^S(\omega)\right]\frac{1-\omega}{\omega}d\omega$. Here we consider an extra assumption $T^2(t)=T_0.T(t)$ [1], where $T=\frac{\alpha_S(t)}{2\pi}$ and $T_0$ is a numerical parameter. By a suitable choice of $T_0$, we can reduce the error to a minimum.

To compare our result with experimental data, we have to consider the relation between proton and deuteron structure functions measured in DIS with non-singlet quark distribution function as $F_2^{NS}(x,t)=3\left(2F_2^p(x,t)-F_2^d(x,t)\right)$.

In this report, we have compared our results of t and x-evolutions of non-singlet structure function $F_2^{NS}$ (x, t) with E665 experiment data [3] (taken at Fermilab in inelastic muon scattering with an average beam energy of 470 GeV) and NMC data [4] (in muon-deuteron DIS with incident momentum 90, 120, 200, 280 GeV). We consider the range $0.01 \le x \le 0.0489$ and $1.496 \le Q^2 \le 13.391$ GeV$^2$ for E665 data and $0.0045 \le x \le 0.14$ and $0.75 \le Q^2 \le 20$ GeV$^2$ for NMC data. It is observed that, within these range, for the lowest error $T_0$ = 0.05 (Fig. 1). Fig. 2(a) and Fig. 3(a) represent the t-evolution and Fig. 2(b) and Fig. 3(b) represent x-evolution of non-singlet structure function. Here errors are statistical and systematic uncertainty. Here in all graphs, lowest-t and highest-x points are taken as inputs $F_2^{NS}$ (x, $t_0$) and $F_2^{NS}$ ($x_0$, t) respectively. For quantitative analysis, we consider the QCD cut off parameter $\Lambda_{\overline{MS}}$ =0.323 GeV [8] and $N_f$ = 4. It is observed that, our results are compatible with experimental values. Again fitness is better in NLO than that of LO.

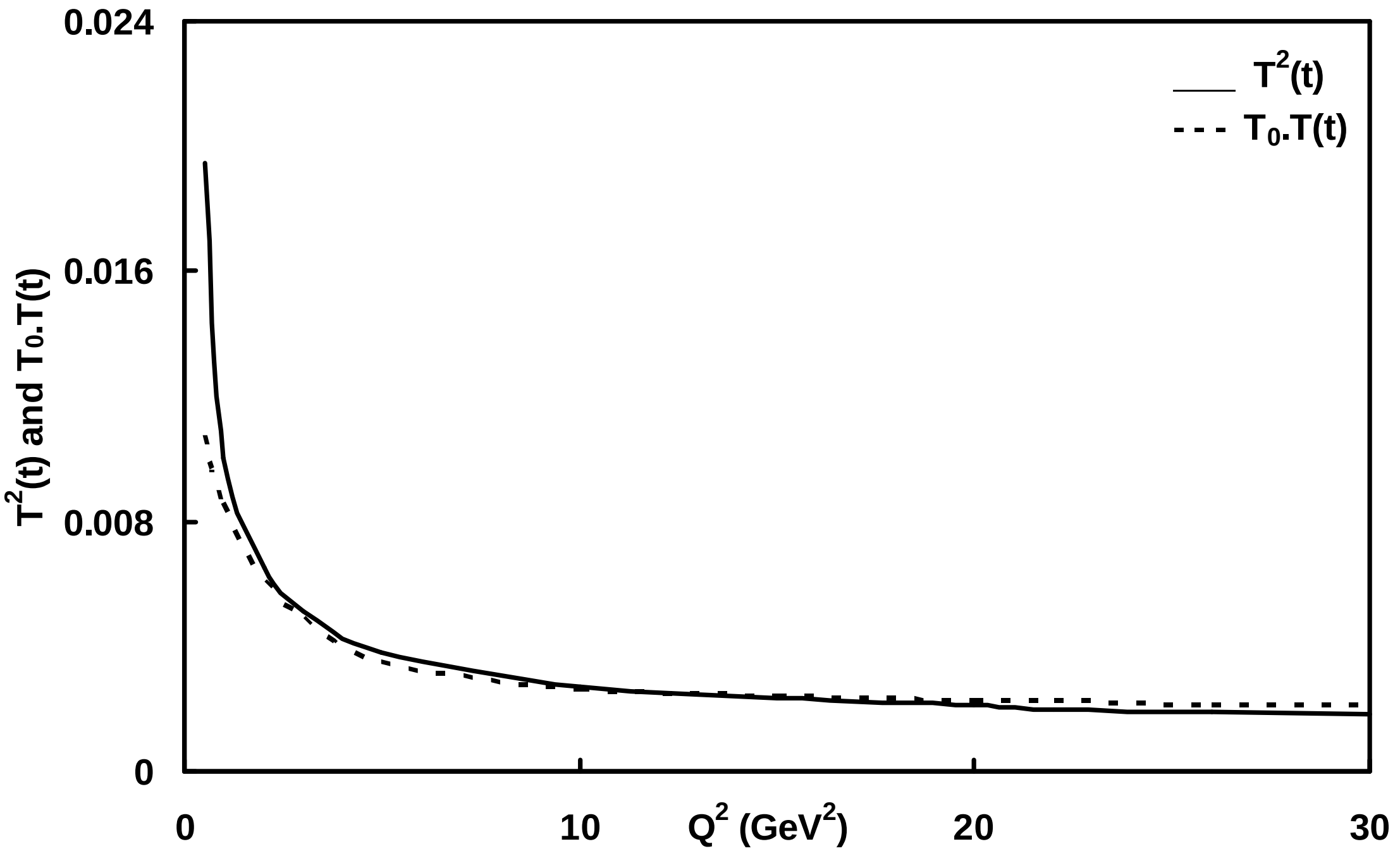

**FIG. 1.** $T^2(t)$ and $T_0.T(t)$ versus $Q^2$ ($GeV^2$)

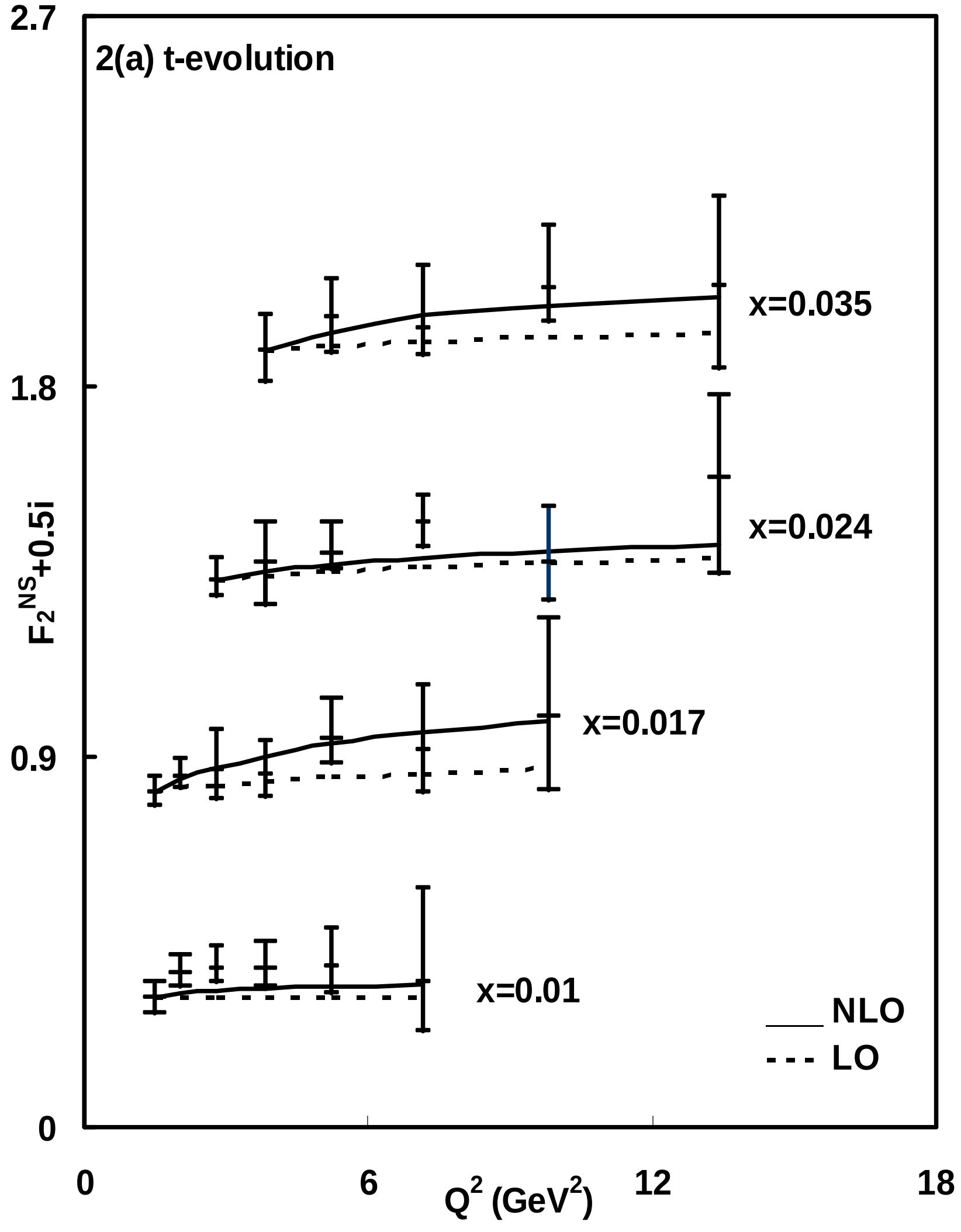

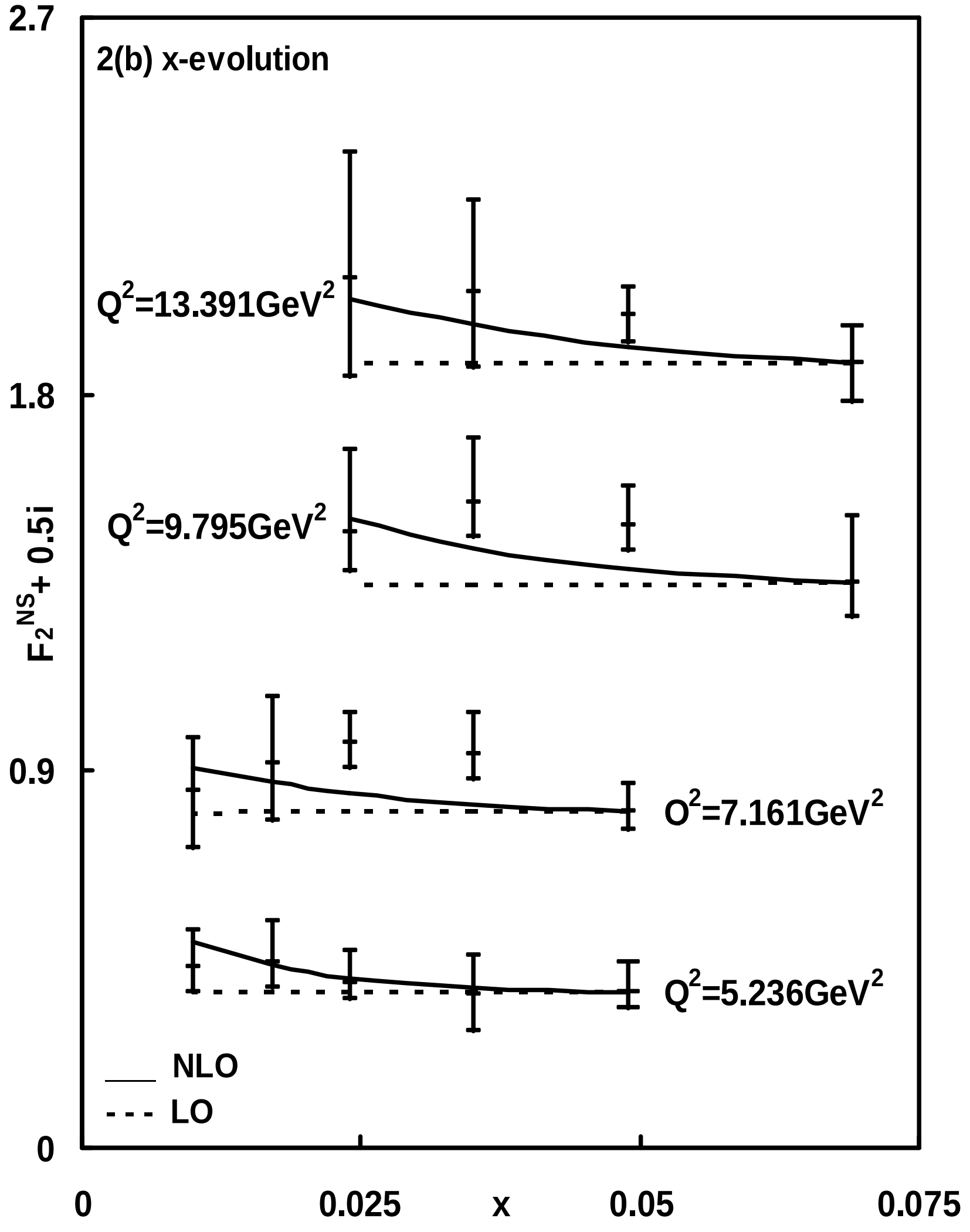

**FIG. 2(a,b): E665 data. Dotted lines are our LO results and solid lines are NLO results. For clarity, data are scaled up by +0.5i (i=0, 1, 2, 3) starting from bottom of all graphs in each figure.**

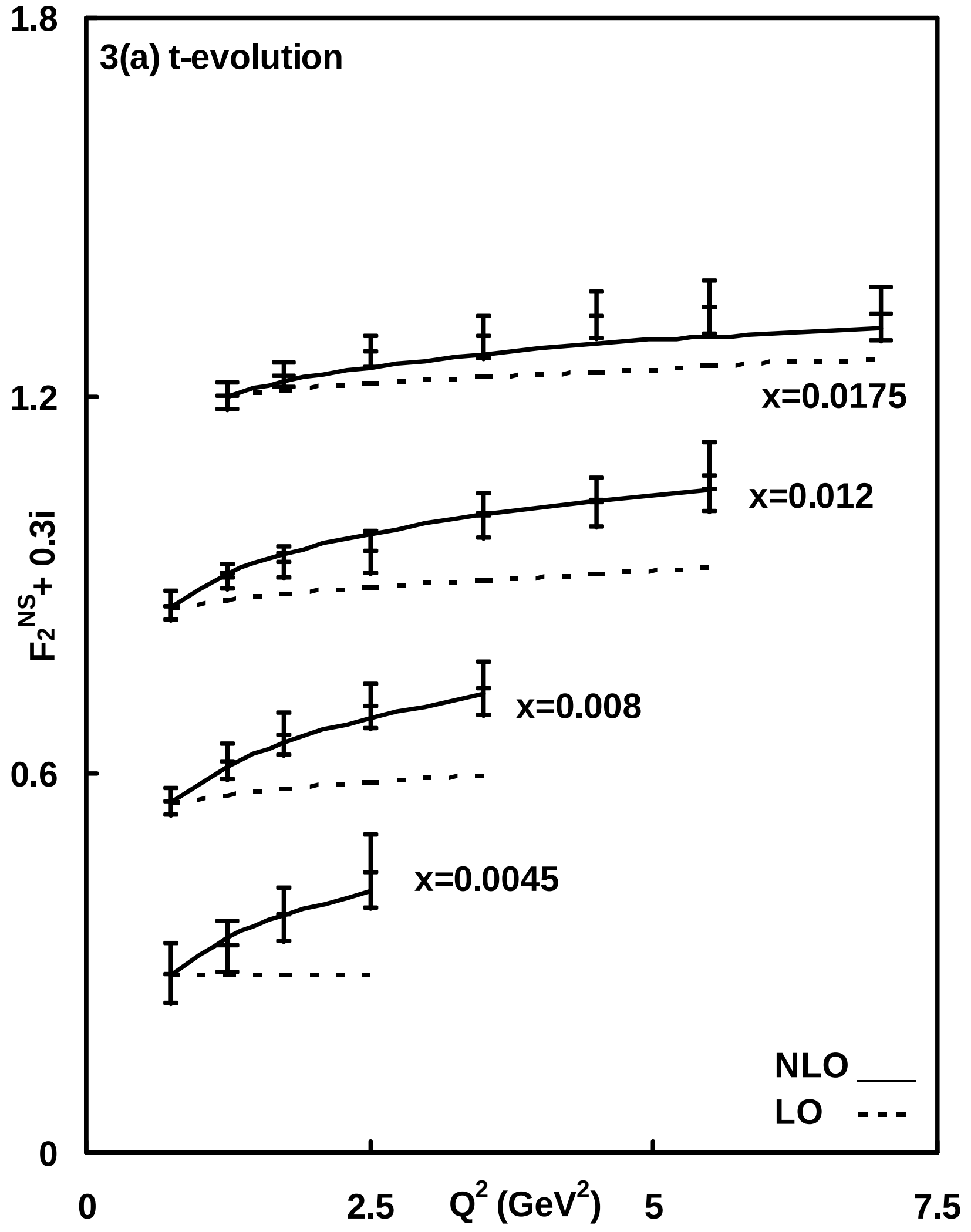

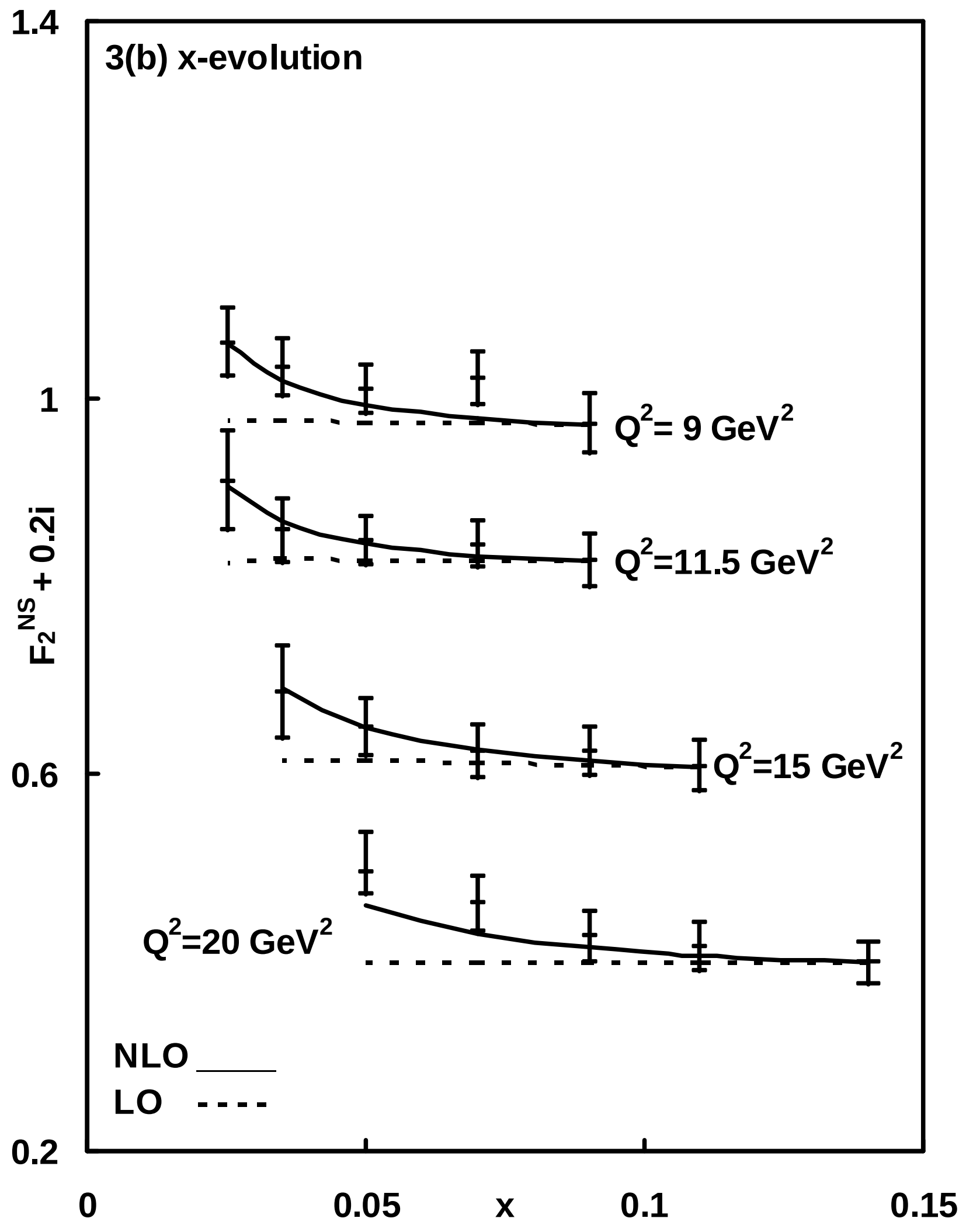

**FIG. 3(a,b): NMC data. Dotted lines are our LO results and solid lines are NLO results. For clarity, data are scaled up by +0.3i (in Fig. 3(a)) and +0.2i (in Fig. 3(b)) (with i=0, 1, 2, 3) starting from bottom of all graphs in each figure.**